\documentstyle[12pt]{article}
\newcommand{\pp}{\makebox{$\pi\pi$}}
\newcommand{\pva}{\makebox{$\vec{p}_a$}}
\newcommand{\pvb}{\makebox{$\vec{p}_b$}}
\newcommand{\ra}{\makebox{$\vec{r}_a$}}
\newcommand{\rb}{\makebox{$\vec{r}_b$}}
\newcommand{\ppp}{\makebox{$\psi_{\pi\pi}$}}
\newcommand{\psa}{\makebox{$\psi_a$}}
\newcommand{\psb}{\makebox{$\psi_b$}}
\newcommand{\pta}{\makebox{$\tilde{\psi}_a$}}
\newcommand{\ptb}{\makebox{$\tilde{\psi}_b$}}
\newcommand{\gsim}
{\mbox{\hspace*{0.1em}\raisebox{.4ex}{$\scriptstyle >$}
\hspace{-0.88em}\raisebox{-.6ex}{$\scriptstyle\sim$}\hspace*{0.05em}}}
\newcommand{\lsim}
{\mbox{\hspace*{0.1em}\raisebox{.4ex}{$\scriptstyle <$}
\hspace{-0.88em}\raisebox{-.6ex}{$\scriptstyle\sim$}\hspace*{0.05em}}}
\newcommand{\sd}{\makebox[0.5ex]{}}
\begin{document}
\begin{center}
{\bf Pion Correlations and Two-Particle Bound States}\\[0.3cm]
M.Shmatikov\\[0.2cm]
{\it Institute for Theoretical Physics II,\\
Ruhr Universit\"at Bochum,
D-44780 Bochum, Germany} \\
and\\
{\it Russian Research Center "Kurchatov Institute",\\ 123182 Moscow, Russia}
\footnote{Permanent address}\\[0.5cm]
{\bf Abstract}\\[0.3cm]
\end{center}
A method to investigate possible loosely bound dibaryon states is suggested.
The momentum dependence of a two-pion correlation function is shown to depend
on the (non)existence of a bound state.\\[0.5cm] 

The only dibaryon state which existence does not cause at present any doubts 
is the deuteron. There is also a plausible candidate suggested for
the explanation of the forward angle cross section of the pionic double
charge exchange \cite{dcx}. It is unnecessary to stress that its
experimental observation became possible because of the extremely small
width of the resonance: $\Gamma \sd\lsim\sd 0.5$~MeV \cite{dcx}. 

Other dibaryon states with much larger width, exemplified by a $\Delta\Delta$ 
bound state \cite{dd}, are also conceivable. A naive estimate of its width
$\Gamma \sd \gsim \sd 2\,\Gamma_{\Delta}\approx 300$~MeV prompts, however, 
that it will hardly manifest itself as a resonance peak.

Here we address the problem of observing possible bound states of 
two-baryon resonances.
To be more specific we consider a pair of two baryon resonances ($B_1$ and 
$B_2$), their interaction being strong enough to produce a (loosely-) bound state. 
Were the bound-state components stable it would, like the deuteron, exist
for an infinitely long time. However, because of the baryons' decay the bound state
itself exists for a finite time. The only signal of a bound state emergence will be 
then the localization of the $B_1$ and $B_2$ particles in some finite spatial region. 

A dominant decay channel of a baryon resonance is usually that with 
the $\pi$-meson emission:
\[
B\rightarrow B^{'} + \pi
\]
Then the $(B_1\,B_2)$ bound state will decay with the emission of two pions:
\begin{equation}
(B_1\,B_2) \rightarrow B^{'}_{1} + B{'}_{2} + \pi + \pi
\label{bb}
\end{equation}
In the case of a $(\Delta\Delta)$ bound state the decay pattern (\ref{bb}) 
translates into
\begin{equation}
(\Delta\Delta)\rightarrow N + N + \pi + \pi
\label{nn}
\end{equation}
Hereafter we shall address specifically the process (\ref{nn}) though all the results are 
applicable to a  more general case (\ref{bb}).

Spatial dis\-tri\-bu\-tion of  two $\Delta$-isobars can be scanned using the 
two-particle correlation in the momentum space. The latter in the form of 
the pionic interferometry is routinely applied for extracting information about 
the space-time structure of the emitting sources in heavy-ion collisions \cite{he}.

Forbidding complexity of the multiparticle process (\ref{bb}) (and (\ref{nn})) 
necessitates making some simplifying approximations.  In the vein of the 
$N_c\rightarrow\infty$ approach \cite{ma} we consider the $\Delta$-isobar (and the 
nucleon) as a heavy source of $\pi$-mesons. Then the process (\ref{nn}) can be viewed 
as an emission of the pions by two heavy sources superimposed by the motion of the
sources which is not affected by the pion emission.

Separating the system under consideration into two components: a fast 
($\Delta\rightarrow N\pi$ decay) and a slow (relative motion of two
$\Delta$-isobars and nucleons) ones we apply the adiabatic approximation
\cite{sc}

Consider first two $\Delta$-isobars at a distance $\vec{R}_{\Delta\Delta}$. 
Correlation between two (identical) pions due to the quantum mechanical
uncertainty depends on the sources' separation $\vec{R}_{\Delta\Delta}$. In what 
follows we use the formalism developed in \cite{co}. Let $\ppp$ be the wave function
of the \pp~system. Its time evolution is governed by the hamiltonian $H$:
\begin{equation}
\ppp(\ra,\rb;t) = \exp\left(-i H(\ra,\rb)(t-t_0)\right)\ppp(\ra,\rb;t_0)
\end{equation}
We define also an asymptotic state with definite values of the $\pi$-meson
momenta
\begin{equation}
\phi_{\vec{p}_a,\vec{p}_b}\left(\ra,\rb;t\right)=
\exp\left(-i(E_a+E_b)t\right)
\exp\left(i(\pva\ra+\pvb\rb)\right)
\end{equation}
where $E_{a,b}=\sqrt{m^2_{\pi}+\vec{p}^{\sd 2}_{a,b}}$ are the particle energies.
Then the measured two-particle amplitude equals \cite{co}:
\begin{equation}
\begin{array}{ll}
A(\pva,\pvb) =& \lim_{t\rightarrow\infty}\int d^3r_a\,d^3r_b\,
\ppp(\ra,\rb,t_0)\\
&\times\left[\exp\left[-iH(\ra,\rb)(t-t_0)\right]
\phi_{\vec{p}_a,\vec{p}_b}\left(\ra,\rb;t\right)\right]^*
\end{array}
\label{am}
\end{equation}
The values of $t$ for which this expression is valid will be discussed below.

It is convenient to introduce  center-of-mass and relative coordinates
of two $\pi$-mesons:
\begin{equation}
\vec{r} = \ra - \rb\,;\qquad\qquad \vec{R}=\frac{\ra + \rb}{2}
\end{equation}
and also the total and the relative momenta of the pair:
\begin{equation}
\vec{q}=\pva - \pvb\,;\qquad\qquad \vec{P} = \pva + \pvb
\end{equation}
Using these coordinates the asymptotic wave function can be rewritten as:
\begin{equation}
\phi_{\vec{p}_a,\vec{p}_b}\left(\ra,\rb;t\right)=
e^{i\left(\vec{P}\vec{R}-Et\right)}
e^{\frac{\vec{q}\vec{r}}{2}}
\end{equation}
where $E = E_a + E_b$ is the total energy of the \pp~pair. The total energy
$E$ may be separated then into the c.m.s motion energy $E_{\rm cm}$ and
the relative motion energy $E_{\rm rel}$. The former obviously reads
$E_{\rm cm} = \sqrt{\vec{P}^2 + 4m^2_{\pi}}$. Then $E_{\rm rel} = 
E - E_{\rm cm}$. In a frame where $\vec{P}\approx 0$ the relative motion
energy reduces to $E_{\rm rel}\approx \vec{q}^{\sd 2}/(4m_{\pi})$. We make a 
simplifying assumption that the emitted $\pi$-mesons do not interact with each other. 
(The validity of this approximation is discussed below.)
Then the two-pion amplitude (\ref{am}) can be cast in the form \cite{co}:
\begin{equation}
A(\pva,\pvb)= e^{iEt_0}\int d^3r\,d^3R\,e^{-i\vec{P}\vec{R}}\,e^{i\vec{q}\vec{r}/2} \,
\ppp\left(\vec{R}+\frac{\vec{r}}{2}, \vec{R} -\frac{\vec{r}}{2};t_0\right)
\end{equation}

We assume that two pions are emitted in the decays of baryons independently, 
implying that at some time moment $t_a$ the two-particle wave function 
\ppp~ factorizes:
\begin{equation}
\ppp(\ra,\rb;t_a)=  \frac{1}{\sqrt{2}}\left[\psi_a(\ra;t_a)\,\psi_b(\rb;t_a)
+ \psi_a(\rb;t_a)\,\psi_b(\ra;t_a)\right]
\end{equation}

Consider the case when the $\pi$-meson, in the state described by the $\psi_b$ wave 
function, was emitted earlier than that in the $\psi_a$ state at some time $t_b < t_a$.
Then by the moment $t_a$ (when the second pion was emitted) its wave function has 
acquired some phase:
\begin{equation}
\psb(\vec{r};t_a)= e^{-i\,h(\vec{r})(t_a-t_b)}\,\psb(\vec{r};t_b)
\label{wf}
\end{equation} 
where $h$ is the one-particle Hamiltonian (of free motion): 
$h(\vec{r}) = m - \frac{1}{2m}\nabla^2_{\vec{r}}$.
The two-pion wave function reads then \cite{co}:
\begin{equation}
\begin{array}{ll}
\ppp(\ra,\rb;t) = \frac{\theta(t_a-t_b)}{\sqrt{2}}&\left[\right.\psa(\ra;t_a)\,
e^{-i\,h(\vec{r}_b)(t_a-t_b)}\,\psb(\rb;t_b)\\
&\left. + \psa(\rb;t_a)\,e^{-i\,h(\vec{r}_a)(t_a-t_b)}\psb(\ra;t_b)\right] 
\end{array}
\end{equation}
The time evolution of the wave function(s) is described the most easily by
introducing the Fourier decomposition into the momentum eigenstates. 
The two-particle amplitude corresponding to the appearance of the 
second $\pi$-meson at the time moment $t_a$ can be represented then as follows 
\cite{co}:
\begin{equation}
\begin{array}{ll}
A^{(a)}_{ab}(\pva,\pvb)= &\sqrt{2}\,\theta(t_a-t_b)\int 
\frac{d^3Q}{(2\pi)^3}\,\frac{d^3k}{(2\pi)^3}\,d^3r\,d^3 R\,e^{i(\vec{Q}-\vec{P})
\cdot\vec{R}}\,
e^{i\Omega(\vec{Q},\vec{k})t_a}\\ 
&\times \cos\left(\frac{\vec{q}\vec{r}}{2}\right)\,
\pta\left(\frac{\vec{Q}}{2}+\vec{k};t_a\right)
\ptb\left(\frac{\vec{Q}}{2}-\vec{k};t_b\right) 
\end{array}
\label{ab}
\end{equation}
Here $\tilde{\psi}(\vec{k};t)$ is the Fourier transform of the one-particle
wave function defined as follows
\begin{equation}
\psi(\vec{r};t) =\int \frac{d^3 k}{(2\pi)^3}\,\tilde{\psi}(\vec{k};t)\,
e^{i(\vec{k}\vec{r}-\omega t)}
\end{equation}
and $\omega=\sqrt{m^2_\pi + \vec{k}^2}$. In eq.(\ref{ab}) 
$\vec{Q}=\vec{k}_a + \vec{k}_b$ and $\vec{k}=(\vec{k}_a - \vec{k}_b)/2$.
Finally, $\Omega(\vec{Q},\vec{k})\equiv E-
\omega\left(\vec{Q}/2+\vec{k}\right) -\omega\left(\vec{Q}/2-\vec{k}\right)$.
The integral over the c.m.s. coordinate $\vec{R}$ yields 
$(2\pi)^3\delta^{(3)}(\vec{Q}-\vec{P})$, while the integration over $\vec{r}$
produces a Fourier image of the $\cos(\vec{q}\vec{r}/2)$ function 
$\Phi(\vec{q}/2,\vec{k})$:
\begin{equation}
\begin {array}{ll}
A^{(a)}_{ab}(\pva,\pvb)=&\theta(t_a-t_b)\int\frac{d^3k}{(2\pi)^3}\,
e^{i\Omega(\vec{P}/2,\vec{k})t_a}\,\Phi(\vec{q}/2,\vec{k})\\
&\times\pta(\vec{P}/2+\vec{k};t_a)\,\ptb(\vec{P}/2-\vec{k};t_b)
\end{array}
\label{ac}
\end{equation}
An alternative situation when the particle in the state $a$ was emitted earlier
at the time $t_a$ and the  other particle appeared later at the time $t_b$
yields similar expression \cite{co}:
\begin{equation}
\begin{array}{ll}
A^{(b)}_{ab}(\pva,\pvb)=&\theta(t_b-t_a)\int\frac{d^3k}{(2\pi)^3}\,
e^{i\Omega(\vec{P}/2,\vec{k})t_b}\,\Phi(\vec{q}/2,\vec{k})\\
&\times\pta(\vec{P}/2+\vec{k};t_a)\,\ptb(\vec{P}/2-\vec{k};t_b)
\end{array}
\label{ad}
\end{equation}
The total amplitude is the sum of two components (eq.(\ref{ac}) and (\ref{ad}))
corresponding to two time orderings of the particles emission:
\begin{equation}
A_{ab}(\pva,\pvb)=A^{(a)}_{ab}(\pva,\pvb) + A^{(b)}_{ab}(\pva,\pvb)
\end{equation}
Using this amplitude we calculate the two-particle cross section which
is defined as follows \cite{co}
\begin{equation}
P_2(\pva,\pvb)=\sum_{ab,a^{'}b^{'}}\int dt_a\,dt_b\,dt_{a^{'}}\,
dt_{b^{'}}\,\rho_{ab,a^{'}b^{'}}\,A^*_{a^{'}b^{'}}(\pva,\pvb)
A_{ab}(\pva,\pvb)
\label{pp}
\end{equation}

Here $\rho_{ab,a^{'}b^{'}}$ is the density matrix of the source which contains 
information on the probability distribution for the two-particle quantum numbers 
$(a,b)$ and for the emission times $(t_a,t_b)$. 

Independent emission of two particles is ensured by a factorizable
ansatz for the density matrix \cite{co}:
\begin{equation}
\rho_{ab,a^{'}b^{'}}=\nu_{aa^{'}}\,\rho(t_a,t_{a^{'}})\,
\nu_{bb^{'}}\,\rho(t_b,t_{b^{'}})
\end{equation}

The probability of the two-particle emission consists of four terms:
\begin{equation}
P_2(\pva,\pvb)= P^{(aa)} + P^{(bb)} + P^{(ab)} + P^{(ba)}
\end{equation}
The reader is referred to \cite{co} for the proof that $P^{(ab)}=P^{(ba)}=0$ 
and $P^{(aa)}=P^{(bb)}$ so that
\begin{equation}
P_2(\pva,\pvb) = 2P^{(aa)}
\end{equation}

Following \cite{co} we define the single particle density $S(X,K)$ of the
source:
\begin{equation}
\begin{array}{ll}
S(X,K) = & \int d^4x\,e^{i(K\cdot x)}\,\rho(X^0+x^0/2, X^0 - x^0/2)\\
& \times \sum_{aa^{'}}\nu_{aa^{'}}\,\psa(X + x/2)\,\psi^{*}_{a^{'}}(X - x/2)
\end{array}
\label{sd}
\end{equation}
where 
\begin{equation}
X^0 =\frac{1}{2}(t_a+t_{a^{'}}), \qquad\qquad x^0 = t_a - t_{a^{'}} 
\end{equation}
and the 4-momentum $K$ equals:
\begin{equation}
\vec{K}=\frac{\pva + \pvb}{2}\,;\qquad\qquad K_0 =\frac{E_a + E_b}{2}
\end{equation}
Then the two-particle probability (\ref{pp}) can be expressed in terms of the 
single-particle density \cite{co}:
\begin{equation}
\begin{array}{ll}
P_2(\pva,\pvb)=& \int d^4X\,d^4Y\,\left[S(X,K +q/2)S(Y,K-q/2)\right.\\ 
& \left. + e^{iq(X-Y)}\,S(X,K) S(Y,K)\right]
\end{array}
\label{ma}
\end{equation}
Here $q_0 = E_a - E_b$ and $\vec{q}= \pva-\pvb$. Remind that the expression
for the two-particle probability was obtained in the frame where
$\pva + \pvb\approx 0$ implying that $q_0 \approx 0$. Single particle spectrum is 
given by the expression:
\begin{equation}
P_1(\vec{p}) = \int d^4X\, S(X,p)
\end{equation}
with the 4-momentum $p=(\sqrt{\vec{p}^{\sd 2} + m^2_\pi}, \vec{p})$

We are now in position to investigate the two-particle correlation
function $C(\pva,\pvb)\equiv C(\vec{q},\vec{K})$:
\begin{equation}
C(\vec{q},\vec{K})=\frac{P_2(\pva,\pvb)}{P_1(\pva)\,P_1(\pvb)}
\label{cc}
\end{equation}

We simplify this expression by separating the time and the spatial dependencies of 
the single-particle density $S(X,K)$ (eq.(\ref{sd})). To this end we introduce a 
short-hand notation
\begin{equation}
\rho(T,t)\equiv \rho(X^0+ x^0/2;X^0 - x^0/2)
\end{equation}
denoting the rest of the $S(X,K)$ as $\tilde{S}$: 
\begin{equation}
\tilde{S}(\vec{R},\vec{K}) = \int d^3r\,e^{i\vec{K}\vec{r}}\sum_{aa^{'}}\nu_{aa{'}}
\psi_a\left(\vec{R}+\vec{r}/2\right)\,\psi^{*}_{a^{'}}\left(\vec{R}-\vec{r}/2\right)
\label{ts}
\end{equation} so that
\begin{equation}
S(X,K) = \int dt\,e^{iEt}\,\rho(T,t)\,\tilde{S}(\vec{R},\vec{K});\,\qquad\qquad X
\equiv(T,\vec{R})
\end{equation}
Then the two-particle probability (\ref{ma}) can be cast in the form:
\begin{equation}
\begin{array}{l}
P_2= 
 \int dT_1\,dT_2\,d^3R_1\,d^3R_2\,\int dt_1\,dt_2
\left\{ e^{iE_at_1}\,e^{iE_bt_2}\,\rho_1\,\rho_2\,
\tilde{S}(\vec{R}_1,\pva)\,\tilde{S}(\vec{R}_2,\pvb)\right.\\ 
 + e^{i(E_a-E_b)(T_1-T_2)}\,e^{i\vec{q}(\vec{R}_1-\vec{R}_2)}
\left. e^{iK_0t_1}\,e^{iK_0t_2}\rho_1\,\rho_2\,
\tilde{S}(\vec{R}_1,\vec{K}_1)\,\tilde{S}(\vec{R}_2,\vec{K}_2)\right\}
\end{array}
\end{equation}
where we introduced for notational brevity $\rho_i\equiv \rho(T_i,t_i);\; i=1,2$.
Note that in the considered frame $\vec{K}\approx0$ the energies of the pions
are almost equal to each other ($E_a\approx E_b$), implying that 
$K_0\approx E_a(\approx E_b)$. The $P_2$ probability may be simplified further by
introducing an (energy-dependent) factor
\begin{equation}
\lambda(E) = \int dT\,dt\,\rho(T,t)\, e^{iEt}
\end{equation}
We arrive at
\begin{equation}
\begin{array}{l}
P_2(\pva,\pvb)= \lambda(E_a)\, \lambda(E_b)\int d^3R_1\,d^3R_2 \\
\left\{\tilde{S}(\vec{R}_1,\pva)\,\tilde{S}(\vec{R}_2,\pvb)  +
e^{i(\vec{p}_a-\vec{p}_b)(\vec{R}_1-\vec{R}_2)}
 \tilde{S}\left(\vec{R}_1,\frac{\vec{p}_a+\vec{p}_b}{2}\right)\,
\tilde{S}\left(\vec{R}_2,\frac{\vec{p}_a+\vec{p}_b}{2}\right)\right\}.
\end{array}
\end{equation}
The single-particle probability in these notations reads:
\begin{equation}
P_1(\pva)=\lambda(E_a)\,\int d^3R\,\tilde{S}(\vec{R},\pva)
\end{equation}
with the similar expression for $P_1(\pvb)$.

We obtain finally the two-particle correlation function:
\begin{equation}
\begin{array}{l}
C(\pva,\pvb) = 1+\\
\int d^3R_1\,d^3R_2\,
e^{(\vec{p}_a-\vec{p}_b)(\vec{R}_1-\vec{R}_2)}\,
\tilde{S}\left(\vec{R}_1, \frac{\vec{p}_a + \vec{p}_b}{2}\right)\,
\tilde{S}\left(\vec{R}_2, \frac{\vec{p}_a + \vec{p}_b}{2}\right) \\
\left\{\int d^3R_1\,\tilde{S}(\vec{R}_1,\pva)\,
\int d^3R_2\,\tilde{S}(\vec{R}_2,\pvb)\right\}^{-1}
\end{array}
\end{equation}

Let us consider now a bound state of two baryons. As it was stated above,
in the $N_c\rightarrow\infty$ limit they can be considered as heavy
sources of $\pi$-mesons. Neglecting recoil due to the pion emission
we can write the spatial distribution of the sources  as
\begin{equation}
\psi(\vec{r}) = \gamma\left\{\delta(\vec{r}-\vec{R}_{\Delta\Delta}/2)+ 
\delta(\vec{r}+\vec{R}_{\Delta\Delta}/2)\right\}
\end{equation}
where $\vec{R}_{\Delta\Delta}$ is the distance between two $\Delta$-isobars (baryons) 
and $\gamma$ is a normalization
factor. (Hereafter for notational brevity we omit the  wave function subscript 
denoting the charge state  of the emitted pion). Inserting the wave function into the 
expression for the $\tilde{S}$ (eq.(\ref{ts})) we arrive at
\begin{equation}
\tilde{S}(\vec{R},\vec{p})=\gamma^2\left\{\delta(\vec{R} - \vec{R}_{\Delta\Delta}/2)+
\delta(\vec{R}+\vec{R}_{\Delta\Delta}/2)\right\}
\label{sm}
\end{equation}
and the corresponding single-particle distribution
\begin{equation}
P_1(\pva) =\lambda(E_a)\,2\gamma^2
\label{sp}
\end{equation}
The two-particle distribution calculated with the same source function (\ref{sm}) 
reads 
\begin{equation}
P_2 (\pva,\pvb) = 2\gamma^4(1 + \cos(\vec{q}\vec{R}_{\Delta\Delta}))
\label{tp}
\end{equation}
where $\vec{q}=\pva-\pvb$. Note that in deducing $P_2$ only those configuration
were taken into consideration when each heavy baryon emits a pion, i.e. the 
$\pi$-mesons are emitted in different spatial points.

Inserting the single- (eq.(\ref{sp})) and two-particle (eq. (\ref{tp})) distributions
in the expression (\ref{cc}) we arrive at the correlation function for pions emitted 
by two heavy sources separated by the distance $\vec{R}$ (since it will not cause any 
confusion the subscript $\Delta\Delta$ is omitted hereafter):
\begin{equation}
C_{\vec{R}} = 1 + \cos(\vec{q}\vec{R})
\label{cr}
\end{equation}
Again, only those configurations were taken into account when each $\Delta$-isobar
de-excitates by the emission of the $\pi$-meson. It implies that the $C_{\vec{R}}$
correlation function is contributed by the configurations when the emission points
of the $\pi$-mesons are spatially separated.
Note also that since $\pva+\pvb\approx 0$, the relative momentum of two pions equals 
$\vec{q}\approx 2\pva$.

The correlation function (\ref{cr}) depends parametrically on the distance between 
pion-emitting sources $\vec{R}$. Two baryons being bound, their spatial distribution 
is governed by the bound-state wave function $\phi(\vec{R})$: the probability that 
the baryons are at the distance $\vec{R}$ is given by the $|\phi(\vec{R})|^2$. 
The spatial distribution of pion-emitting sources does not affect the single-particle
distribution (\ref{sp}), whereas the two-particle distribution (\ref{tp}) and, 
consequently, the two-pion correlation function $C_{\vec{R}}$ is modified. The 
observable correlation function
\begin{equation}
C= \int d^3R\,C_{\vec{R}}\,|\phi(\vec{R})|^2
\end{equation}
reads
\begin{equation}
C(\pva,\pvb) = 1 + \int d^3R\,\cos(\vec{q}\vec{R})\,|\phi(\vec{R})|^2
\label{cf}
\end{equation}
This expression shows that the two-pion correlation function with $\vec{q}$ varying 
"scans" the wave function of the baryon relative motion. 

Eq. (\ref{cf}) becomes more "transparent" in the case of an $S$-wave two-baryon bound 
state. Assuming a simplest deuteron-like wave function
\begin{equation}
\phi(\vec{R})= \sqrt{\frac{\kappa}{2\pi}}\,\frac{e^{-\kappa\,R}}{R}
\end{equation}
where $\kappa$ is the bound-state momentum ($\kappa=\sqrt{\varepsilon\,M}$, 
$\varepsilon$ is the binding energy and $M$ is the reduced mass of the pair of 
baryons) after straightforward integration over angular variables in eq.(\ref{cf}) 
we arrive at
\begin{equation}
C = 1 + 2\kappa\,\int e
^{-2\kappa R}\,\frac{\sin(qR)}{qR}\,dR
\end{equation} 
Introducing the $\chi = q/2\kappa$ ratio and calculating the integral we obtain the 
two-pion correlation function reading:
\begin{equation}
C=1 + \frac{\arctan\chi}{\chi}
\label{ch}
\end{equation}
The dependence of the correlation function $C$ on $\chi$ is plotted in the figure.
\begin{figure}
\setlength{\unitlength}{0.240900pt}
\ifx\plotpoint\undefined\newsavebox{\plotpoint}\fi
\sbox{\plotpoint}{\rule[-0.200pt]{0.400pt}{0.400pt}}%
\begin{picture}(1500,900)(0,0)
\font\gnuplot=cmr10 at 10pt
\gnuplot
\sbox{\plotpoint}{\rule[-0.200pt]{0.400pt}{0.400pt}}%
\put(220.0,113.0){\rule[-0.200pt]{292.934pt}{0.400pt}}
\put(220.0,113.0){\rule[-0.200pt]{0.400pt}{184.048pt}}
\put(220.0,113.0){\rule[-0.200pt]{4.818pt}{0.400pt}}
\put(198,113){\makebox(0,0)[r]{0}}
\put(1416.0,113.0){\rule[-0.200pt]{4.818pt}{0.400pt}}
\put(220.0,266.0){\rule[-0.200pt]{4.818pt}{0.400pt}}
\put(198,266){\makebox(0,0)[r]{0.5}}
\put(1416.0,266.0){\rule[-0.200pt]{4.818pt}{0.400pt}}
\put(220.0,419.0){\rule[-0.200pt]{4.818pt}{0.400pt}}
\put(198,419){\makebox(0,0)[r]{1}}
\put(1416.0,419.0){\rule[-0.200pt]{4.818pt}{0.400pt}}
\put(220.0,571.0){\rule[-0.200pt]{4.818pt}{0.400pt}}
\put(198,571){\makebox(0,0)[r]{1.5}}
\put(1416.0,571.0){\rule[-0.200pt]{4.818pt}{0.400pt}}
\put(220.0,724.0){\rule[-0.200pt]{4.818pt}{0.400pt}}
\put(198,724){\makebox(0,0)[r]{2}}
\put(1416.0,724.0){\rule[-0.200pt]{4.818pt}{0.400pt}}
\put(220.0,877.0){\rule[-0.200pt]{4.818pt}{0.400pt}}
\put(198,877){\makebox(0,0)[r]{2.5}}
\put(1416.0,877.0){\rule[-0.200pt]{4.818pt}{0.400pt}}
\put(220.0,113.0){\rule[-0.200pt]{0.400pt}{4.818pt}}
\put(220,68){\makebox(0,0){0}}
\put(220.0,857.0){\rule[-0.200pt]{0.400pt}{4.818pt}}
\put(341.0,113.0){\rule[-0.200pt]{0.400pt}{4.818pt}}
\put(341,68){\makebox(0,0){0.5}}
\put(341.0,857.0){\rule[-0.200pt]{0.400pt}{4.818pt}}
\put(463.0,113.0){\rule[-0.200pt]{0.400pt}{4.818pt}}
\put(463,68){\makebox(0,0){1}}
\put(463.0,857.0){\rule[-0.200pt]{0.400pt}{4.818pt}}
\put(585.0,113.0){\rule[-0.200pt]{0.400pt}{4.818pt}}
\put(585,68){\makebox(0,0){1.5}}
\put(585.0,857.0){\rule[-0.200pt]{0.400pt}{4.818pt}}
\put(706.0,113.0){\rule[-0.200pt]{0.400pt}{4.818pt}}
\put(706,68){\makebox(0,0){2}}
\put(706.0,857.0){\rule[-0.200pt]{0.400pt}{4.818pt}}
\put(828.0,113.0){\rule[-0.200pt]{0.400pt}{4.818pt}}
\put(828,68){\makebox(0,0){2.5}}
\put(828.0,857.0){\rule[-0.200pt]{0.400pt}{4.818pt}}
\put(950.0,113.0){\rule[-0.200pt]{0.400pt}{4.818pt}}
\put(950,68){\makebox(0,0){3}}
\put(950.0,857.0){\rule[-0.200pt]{0.400pt}{4.818pt}}
\put(1071.0,113.0){\rule[-0.200pt]{0.400pt}{4.818pt}}
\put(1071,68){\makebox(0,0){3.5}}
\put(1071.0,857.0){\rule[-0.200pt]{0.400pt}{4.818pt}}
\put(1193.0,113.0){\rule[-0.200pt]{0.400pt}{4.818pt}}
\put(1193,68){\makebox(0,0){4}}
\put(1193.0,857.0){\rule[-0.200pt]{0.400pt}{4.818pt}}
\put(1314.0,113.0){\rule[-0.200pt]{0.400pt}{4.818pt}}
\put(1314,68){\makebox(0,0){4.5}}
\put(1314.0,857.0){\rule[-0.200pt]{0.400pt}{4.818pt}}
\put(1436.0,113.0){\rule[-0.200pt]{0.400pt}{4.818pt}}
\put(1436,68){\makebox(0,0){5}}
\put(1436.0,857.0){\rule[-0.200pt]{0.400pt}{4.818pt}}
\put(220.0,113.0){\rule[-0.200pt]{292.934pt}{0.400pt}}
\put(1436.0,113.0){\rule[-0.200pt]{0.400pt}{184.048pt}}
\put(220.0,877.0){\rule[-0.200pt]{292.934pt}{0.400pt}}
\put(45,495){\makebox(0,0){$C$}}
\put(828,23){\makebox(0,0){$\chi$}}
\put(220.0,113.0){\rule[-0.200pt]{0.400pt}{184.048pt}}
\put(220,724){\usebox{\plotpoint}}
\put(232,722.67){\rule{3.132pt}{0.400pt}}
\multiput(232.00,723.17)(6.500,-1.000){2}{\rule{1.566pt}{0.400pt}}
\put(245,721.67){\rule{2.891pt}{0.400pt}}
\multiput(245.00,722.17)(6.000,-1.000){2}{\rule{1.445pt}{0.400pt}}
\put(257,720.17){\rule{2.500pt}{0.400pt}}
\multiput(257.00,721.17)(6.811,-2.000){2}{\rule{1.250pt}{0.400pt}}
\put(269,718.17){\rule{2.500pt}{0.400pt}}
\multiput(269.00,719.17)(6.811,-2.000){2}{\rule{1.250pt}{0.400pt}}
\multiput(281.00,716.95)(2.695,-0.447){3}{\rule{1.833pt}{0.108pt}}
\multiput(281.00,717.17)(9.195,-3.000){2}{\rule{0.917pt}{0.400pt}}
\multiput(294.00,713.95)(2.472,-0.447){3}{\rule{1.700pt}{0.108pt}}
\multiput(294.00,714.17)(8.472,-3.000){2}{\rule{0.850pt}{0.400pt}}
\multiput(306.00,710.95)(2.472,-0.447){3}{\rule{1.700pt}{0.108pt}}
\multiput(306.00,711.17)(8.472,-3.000){2}{\rule{0.850pt}{0.400pt}}
\multiput(318.00,707.94)(1.797,-0.468){5}{\rule{1.400pt}{0.113pt}}
\multiput(318.00,708.17)(10.094,-4.000){2}{\rule{0.700pt}{0.400pt}}
\multiput(331.00,703.95)(2.472,-0.447){3}{\rule{1.700pt}{0.108pt}}
\multiput(331.00,704.17)(8.472,-3.000){2}{\rule{0.850pt}{0.400pt}}
\multiput(343.00,700.93)(1.267,-0.477){7}{\rule{1.060pt}{0.115pt}}
\multiput(343.00,701.17)(9.800,-5.000){2}{\rule{0.530pt}{0.400pt}}
\multiput(355.00,695.94)(1.651,-0.468){5}{\rule{1.300pt}{0.113pt}}
\multiput(355.00,696.17)(9.302,-4.000){2}{\rule{0.650pt}{0.400pt}}
\multiput(367.00,691.94)(1.797,-0.468){5}{\rule{1.400pt}{0.113pt}}
\multiput(367.00,692.17)(10.094,-4.000){2}{\rule{0.700pt}{0.400pt}}
\multiput(380.00,687.94)(1.651,-0.468){5}{\rule{1.300pt}{0.113pt}}
\multiput(380.00,688.17)(9.302,-4.000){2}{\rule{0.650pt}{0.400pt}}
\multiput(392.00,683.93)(1.267,-0.477){7}{\rule{1.060pt}{0.115pt}}
\multiput(392.00,684.17)(9.800,-5.000){2}{\rule{0.530pt}{0.400pt}}
\multiput(404.00,678.94)(1.797,-0.468){5}{\rule{1.400pt}{0.113pt}}
\multiput(404.00,679.17)(10.094,-4.000){2}{\rule{0.700pt}{0.400pt}}
\multiput(417.00,674.93)(1.267,-0.477){7}{\rule{1.060pt}{0.115pt}}
\multiput(417.00,675.17)(9.800,-5.000){2}{\rule{0.530pt}{0.400pt}}
\multiput(429.00,669.94)(1.651,-0.468){5}{\rule{1.300pt}{0.113pt}}
\multiput(429.00,670.17)(9.302,-4.000){2}{\rule{0.650pt}{0.400pt}}
\multiput(441.00,665.93)(1.267,-0.477){7}{\rule{1.060pt}{0.115pt}}
\multiput(441.00,666.17)(9.800,-5.000){2}{\rule{0.530pt}{0.400pt}}
\multiput(453.00,660.94)(1.797,-0.468){5}{\rule{1.400pt}{0.113pt}}
\multiput(453.00,661.17)(10.094,-4.000){2}{\rule{0.700pt}{0.400pt}}
\multiput(466.00,656.93)(1.267,-0.477){7}{\rule{1.060pt}{0.115pt}}
\multiput(466.00,657.17)(9.800,-5.000){2}{\rule{0.530pt}{0.400pt}}
\multiput(478.00,651.94)(1.651,-0.468){5}{\rule{1.300pt}{0.113pt}}
\multiput(478.00,652.17)(9.302,-4.000){2}{\rule{0.650pt}{0.400pt}}
\multiput(490.00,647.94)(1.797,-0.468){5}{\rule{1.400pt}{0.113pt}}
\multiput(490.00,648.17)(10.094,-4.000){2}{\rule{0.700pt}{0.400pt}}
\multiput(503.00,643.94)(1.651,-0.468){5}{\rule{1.300pt}{0.113pt}}
\multiput(503.00,644.17)(9.302,-4.000){2}{\rule{0.650pt}{0.400pt}}
\multiput(515.00,639.94)(1.651,-0.468){5}{\rule{1.300pt}{0.113pt}}
\multiput(515.00,640.17)(9.302,-4.000){2}{\rule{0.650pt}{0.400pt}}
\multiput(527.00,635.94)(1.651,-0.468){5}{\rule{1.300pt}{0.113pt}}
\multiput(527.00,636.17)(9.302,-4.000){2}{\rule{0.650pt}{0.400pt}}
\multiput(539.00,631.94)(1.797,-0.468){5}{\rule{1.400pt}{0.113pt}}
\multiput(539.00,632.17)(10.094,-4.000){2}{\rule{0.700pt}{0.400pt}}
\multiput(552.00,627.94)(1.651,-0.468){5}{\rule{1.300pt}{0.113pt}}
\multiput(552.00,628.17)(9.302,-4.000){2}{\rule{0.650pt}{0.400pt}}
\multiput(564.00,623.94)(1.651,-0.468){5}{\rule{1.300pt}{0.113pt}}
\multiput(564.00,624.17)(9.302,-4.000){2}{\rule{0.650pt}{0.400pt}}
\multiput(576.00,619.95)(2.472,-0.447){3}{\rule{1.700pt}{0.108pt}}
\multiput(576.00,620.17)(8.472,-3.000){2}{\rule{0.850pt}{0.400pt}}
\multiput(588.00,616.94)(1.797,-0.468){5}{\rule{1.400pt}{0.113pt}}
\multiput(588.00,617.17)(10.094,-4.000){2}{\rule{0.700pt}{0.400pt}}
\multiput(601.00,612.95)(2.472,-0.447){3}{\rule{1.700pt}{0.108pt}}
\multiput(601.00,613.17)(8.472,-3.000){2}{\rule{0.850pt}{0.400pt}}
\multiput(613.00,609.94)(1.651,-0.468){5}{\rule{1.300pt}{0.113pt}}
\multiput(613.00,610.17)(9.302,-4.000){2}{\rule{0.650pt}{0.400pt}}
\multiput(625.00,605.95)(2.695,-0.447){3}{\rule{1.833pt}{0.108pt}}
\multiput(625.00,606.17)(9.195,-3.000){2}{\rule{0.917pt}{0.400pt}}
\multiput(638.00,602.95)(2.472,-0.447){3}{\rule{1.700pt}{0.108pt}}
\multiput(638.00,603.17)(8.472,-3.000){2}{\rule{0.850pt}{0.400pt}}
\multiput(650.00,599.95)(2.472,-0.447){3}{\rule{1.700pt}{0.108pt}}
\multiput(650.00,600.17)(8.472,-3.000){2}{\rule{0.850pt}{0.400pt}}
\multiput(662.00,596.95)(2.472,-0.447){3}{\rule{1.700pt}{0.108pt}}
\multiput(662.00,597.17)(8.472,-3.000){2}{\rule{0.850pt}{0.400pt}}
\multiput(674.00,593.95)(2.695,-0.447){3}{\rule{1.833pt}{0.108pt}}
\multiput(674.00,594.17)(9.195,-3.000){2}{\rule{0.917pt}{0.400pt}}
\multiput(687.00,590.95)(2.472,-0.447){3}{\rule{1.700pt}{0.108pt}}
\multiput(687.00,591.17)(8.472,-3.000){2}{\rule{0.850pt}{0.400pt}}
\put(699,587.17){\rule{2.500pt}{0.400pt}}
\multiput(699.00,588.17)(6.811,-2.000){2}{\rule{1.250pt}{0.400pt}}
\multiput(711.00,585.95)(2.695,-0.447){3}{\rule{1.833pt}{0.108pt}}
\multiput(711.00,586.17)(9.195,-3.000){2}{\rule{0.917pt}{0.400pt}}
\multiput(724.00,582.95)(2.472,-0.447){3}{\rule{1.700pt}{0.108pt}}
\multiput(724.00,583.17)(8.472,-3.000){2}{\rule{0.850pt}{0.400pt}}
\put(736,579.17){\rule{2.500pt}{0.400pt}}
\multiput(736.00,580.17)(6.811,-2.000){2}{\rule{1.250pt}{0.400pt}}
\multiput(748.00,577.95)(2.472,-0.447){3}{\rule{1.700pt}{0.108pt}}
\multiput(748.00,578.17)(8.472,-3.000){2}{\rule{0.850pt}{0.400pt}}
\put(760,574.17){\rule{2.700pt}{0.400pt}}
\multiput(760.00,575.17)(7.396,-2.000){2}{\rule{1.350pt}{0.400pt}}
\put(773,572.17){\rule{2.500pt}{0.400pt}}
\multiput(773.00,573.17)(6.811,-2.000){2}{\rule{1.250pt}{0.400pt}}
\multiput(785.00,570.95)(2.472,-0.447){3}{\rule{1.700pt}{0.108pt}}
\multiput(785.00,571.17)(8.472,-3.000){2}{\rule{0.850pt}{0.400pt}}
\put(797,567.17){\rule{2.700pt}{0.400pt}}
\multiput(797.00,568.17)(7.396,-2.000){2}{\rule{1.350pt}{0.400pt}}
\put(810,565.17){\rule{2.500pt}{0.400pt}}
\multiput(810.00,566.17)(6.811,-2.000){2}{\rule{1.250pt}{0.400pt}}
\put(822,563.17){\rule{2.500pt}{0.400pt}}
\multiput(822.00,564.17)(6.811,-2.000){2}{\rule{1.250pt}{0.400pt}}
\put(834,561.17){\rule{2.500pt}{0.400pt}}
\multiput(834.00,562.17)(6.811,-2.000){2}{\rule{1.250pt}{0.400pt}}
\put(846,559.17){\rule{2.700pt}{0.400pt}}
\multiput(846.00,560.17)(7.396,-2.000){2}{\rule{1.350pt}{0.400pt}}
\put(859,557.17){\rule{2.500pt}{0.400pt}}
\multiput(859.00,558.17)(6.811,-2.000){2}{\rule{1.250pt}{0.400pt}}
\put(871,555.17){\rule{2.500pt}{0.400pt}}
\multiput(871.00,556.17)(6.811,-2.000){2}{\rule{1.250pt}{0.400pt}}
\put(883,553.17){\rule{2.700pt}{0.400pt}}
\multiput(883.00,554.17)(7.396,-2.000){2}{\rule{1.350pt}{0.400pt}}
\put(896,551.67){\rule{2.891pt}{0.400pt}}
\multiput(896.00,552.17)(6.000,-1.000){2}{\rule{1.445pt}{0.400pt}}
\put(908,550.17){\rule{2.500pt}{0.400pt}}
\multiput(908.00,551.17)(6.811,-2.000){2}{\rule{1.250pt}{0.400pt}}
\put(920,548.17){\rule{2.500pt}{0.400pt}}
\multiput(920.00,549.17)(6.811,-2.000){2}{\rule{1.250pt}{0.400pt}}
\put(932,546.17){\rule{2.700pt}{0.400pt}}
\multiput(932.00,547.17)(7.396,-2.000){2}{\rule{1.350pt}{0.400pt}}
\put(945,544.67){\rule{2.891pt}{0.400pt}}
\multiput(945.00,545.17)(6.000,-1.000){2}{\rule{1.445pt}{0.400pt}}
\put(957,543.17){\rule{2.500pt}{0.400pt}}
\multiput(957.00,544.17)(6.811,-2.000){2}{\rule{1.250pt}{0.400pt}}
\put(969,541.67){\rule{3.132pt}{0.400pt}}
\multiput(969.00,542.17)(6.500,-1.000){2}{\rule{1.566pt}{0.400pt}}
\put(982,540.17){\rule{2.500pt}{0.400pt}}
\multiput(982.00,541.17)(6.811,-2.000){2}{\rule{1.250pt}{0.400pt}}
\put(994,538.67){\rule{2.891pt}{0.400pt}}
\multiput(994.00,539.17)(6.000,-1.000){2}{\rule{1.445pt}{0.400pt}}
\put(1006,537.17){\rule{2.500pt}{0.400pt}}
\multiput(1006.00,538.17)(6.811,-2.000){2}{\rule{1.250pt}{0.400pt}}
\put(1018,535.67){\rule{3.132pt}{0.400pt}}
\multiput(1018.00,536.17)(6.500,-1.000){2}{\rule{1.566pt}{0.400pt}}
\put(1031,534.67){\rule{2.891pt}{0.400pt}}
\multiput(1031.00,535.17)(6.000,-1.000){2}{\rule{1.445pt}{0.400pt}}
\put(1043,533.17){\rule{2.500pt}{0.400pt}}
\multiput(1043.00,534.17)(6.811,-2.000){2}{\rule{1.250pt}{0.400pt}}
\put(1055,531.67){\rule{3.132pt}{0.400pt}}
\multiput(1055.00,532.17)(6.500,-1.000){2}{\rule{1.566pt}{0.400pt}}
\put(1068,530.67){\rule{2.891pt}{0.400pt}}
\multiput(1068.00,531.17)(6.000,-1.000){2}{\rule{1.445pt}{0.400pt}}
\put(1080,529.17){\rule{2.500pt}{0.400pt}}
\multiput(1080.00,530.17)(6.811,-2.000){2}{\rule{1.250pt}{0.400pt}}
\put(1092,527.67){\rule{2.891pt}{0.400pt}}
\multiput(1092.00,528.17)(6.000,-1.000){2}{\rule{1.445pt}{0.400pt}}
\put(1104,526.67){\rule{3.132pt}{0.400pt}}
\multiput(1104.00,527.17)(6.500,-1.000){2}{\rule{1.566pt}{0.400pt}}
\put(1117,525.67){\rule{2.891pt}{0.400pt}}
\multiput(1117.00,526.17)(6.000,-1.000){2}{\rule{1.445pt}{0.400pt}}
\put(1129,524.17){\rule{2.500pt}{0.400pt}}
\multiput(1129.00,525.17)(6.811,-2.000){2}{\rule{1.250pt}{0.400pt}}
\put(1141,522.67){\rule{2.891pt}{0.400pt}}
\multiput(1141.00,523.17)(6.000,-1.000){2}{\rule{1.445pt}{0.400pt}}
\put(1153,521.67){\rule{3.132pt}{0.400pt}}
\multiput(1153.00,522.17)(6.500,-1.000){2}{\rule{1.566pt}{0.400pt}}
\put(1166,520.67){\rule{2.891pt}{0.400pt}}
\multiput(1166.00,521.17)(6.000,-1.000){2}{\rule{1.445pt}{0.400pt}}
\put(1178,519.67){\rule{2.891pt}{0.400pt}}
\multiput(1178.00,520.17)(6.000,-1.000){2}{\rule{1.445pt}{0.400pt}}
\put(1190,518.67){\rule{3.132pt}{0.400pt}}
\multiput(1190.00,519.17)(6.500,-1.000){2}{\rule{1.566pt}{0.400pt}}
\put(1203,517.67){\rule{2.891pt}{0.400pt}}
\multiput(1203.00,518.17)(6.000,-1.000){2}{\rule{1.445pt}{0.400pt}}
\put(1215,516.67){\rule{2.891pt}{0.400pt}}
\multiput(1215.00,517.17)(6.000,-1.000){2}{\rule{1.445pt}{0.400pt}}
\put(1227,515.67){\rule{2.891pt}{0.400pt}}
\multiput(1227.00,516.17)(6.000,-1.000){2}{\rule{1.445pt}{0.400pt}}
\put(1239,514.67){\rule{3.132pt}{0.400pt}}
\multiput(1239.00,515.17)(6.500,-1.000){2}{\rule{1.566pt}{0.400pt}}
\put(1252,513.67){\rule{2.891pt}{0.400pt}}
\multiput(1252.00,514.17)(6.000,-1.000){2}{\rule{1.445pt}{0.400pt}}
\put(1264,512.67){\rule{2.891pt}{0.400pt}}
\multiput(1264.00,513.17)(6.000,-1.000){2}{\rule{1.445pt}{0.400pt}}
\put(1276,511.67){\rule{3.132pt}{0.400pt}}
\multiput(1276.00,512.17)(6.500,-1.000){2}{\rule{1.566pt}{0.400pt}}
\put(1289,510.67){\rule{2.891pt}{0.400pt}}
\multiput(1289.00,511.17)(6.000,-1.000){2}{\rule{1.445pt}{0.400pt}}
\put(220.0,724.0){\rule[-0.200pt]{2.891pt}{0.400pt}}
\put(1313,509.67){\rule{2.891pt}{0.400pt}}
\multiput(1313.00,510.17)(6.000,-1.000){2}{\rule{1.445pt}{0.400pt}}
\put(1325,508.67){\rule{3.132pt}{0.400pt}}
\multiput(1325.00,509.17)(6.500,-1.000){2}{\rule{1.566pt}{0.400pt}}
\put(1338,507.67){\rule{2.891pt}{0.400pt}}
\multiput(1338.00,508.17)(6.000,-1.000){2}{\rule{1.445pt}{0.400pt}}
\put(1350,506.67){\rule{2.891pt}{0.400pt}}
\multiput(1350.00,507.17)(6.000,-1.000){2}{\rule{1.445pt}{0.400pt}}
\put(1362,505.67){\rule{3.132pt}{0.400pt}}
\multiput(1362.00,506.17)(6.500,-1.000){2}{\rule{1.566pt}{0.400pt}}
\put(1301.0,511.0){\rule[-0.200pt]{2.891pt}{0.400pt}}
\put(1387,504.67){\rule{2.891pt}{0.400pt}}
\multiput(1387.00,505.17)(6.000,-1.000){2}{\rule{1.445pt}{0.400pt}}
\put(1399,503.67){\rule{2.891pt}{0.400pt}}
\multiput(1399.00,504.17)(6.000,-1.000){2}{\rule{1.445pt}{0.400pt}}
\put(1411,502.67){\rule{3.132pt}{0.400pt}}
\multiput(1411.00,503.17)(6.500,-1.000){2}{\rule{1.566pt}{0.400pt}}
\put(1375.0,506.0){\rule[-0.200pt]{2.891pt}{0.400pt}}
\put(1424.0,503.0){\rule[-0.200pt]{2.891pt}{0.400pt}}
\sbox{\plotpoint}{\rule[-0.500pt]{1.000pt}{1.000pt}}%
\put(220,419){\usebox{\plotpoint}}
\put(220.00,419.00){\usebox{\plotpoint}}
\put(240.76,419.00){\usebox{\plotpoint}}
\multiput(245,419)(20.756,0.000){0}{\usebox{\plotpoint}}
\put(261.51,419.00){\usebox{\plotpoint}}
\multiput(269,419)(20.756,0.000){0}{\usebox{\plotpoint}}
\put(282.27,419.00){\usebox{\plotpoint}}
\put(303.02,419.00){\usebox{\plotpoint}}
\multiput(306,419)(20.756,0.000){0}{\usebox{\plotpoint}}
\put(323.78,419.00){\usebox{\plotpoint}}
\multiput(331,419)(20.756,0.000){0}{\usebox{\plotpoint}}
\put(344.53,419.00){\usebox{\plotpoint}}
\put(365.29,419.00){\usebox{\plotpoint}}
\multiput(367,419)(20.756,0.000){0}{\usebox{\plotpoint}}
\put(386.04,419.00){\usebox{\plotpoint}}
\multiput(392,419)(20.756,0.000){0}{\usebox{\plotpoint}}
\put(406.80,419.00){\usebox{\plotpoint}}
\put(427.55,419.00){\usebox{\plotpoint}}
\multiput(429,419)(20.756,0.000){0}{\usebox{\plotpoint}}
\put(448.31,419.00){\usebox{\plotpoint}}
\multiput(453,419)(20.756,0.000){0}{\usebox{\plotpoint}}
\put(469.07,419.00){\usebox{\plotpoint}}
\put(489.82,419.00){\usebox{\plotpoint}}
\multiput(490,419)(20.756,0.000){0}{\usebox{\plotpoint}}
\put(510.58,419.00){\usebox{\plotpoint}}
\multiput(515,419)(20.756,0.000){0}{\usebox{\plotpoint}}
\put(531.33,419.00){\usebox{\plotpoint}}
\multiput(539,419)(20.756,0.000){0}{\usebox{\plotpoint}}
\put(552.09,419.00){\usebox{\plotpoint}}
\put(572.84,419.00){\usebox{\plotpoint}}
\multiput(576,419)(20.756,0.000){0}{\usebox{\plotpoint}}
\put(593.60,419.00){\usebox{\plotpoint}}
\multiput(601,419)(20.756,0.000){0}{\usebox{\plotpoint}}
\put(614.35,419.00){\usebox{\plotpoint}}
\put(635.11,419.00){\usebox{\plotpoint}}
\multiput(638,419)(20.756,0.000){0}{\usebox{\plotpoint}}
\put(655.87,419.00){\usebox{\plotpoint}}
\multiput(662,419)(20.756,0.000){0}{\usebox{\plotpoint}}
\put(676.62,419.00){\usebox{\plotpoint}}
\put(697.38,419.00){\usebox{\plotpoint}}
\multiput(699,419)(20.756,0.000){0}{\usebox{\plotpoint}}
\put(718.13,419.00){\usebox{\plotpoint}}
\multiput(724,419)(20.756,0.000){0}{\usebox{\plotpoint}}
\put(738.89,419.00){\usebox{\plotpoint}}
\put(759.64,419.00){\usebox{\plotpoint}}
\multiput(760,419)(20.756,0.000){0}{\usebox{\plotpoint}}
\put(780.40,419.00){\usebox{\plotpoint}}
\multiput(785,419)(20.756,0.000){0}{\usebox{\plotpoint}}
\put(801.15,419.00){\usebox{\plotpoint}}
\put(821.91,419.00){\usebox{\plotpoint}}
\multiput(822,419)(20.756,0.000){0}{\usebox{\plotpoint}}
\put(842.66,419.00){\usebox{\plotpoint}}
\multiput(846,419)(20.756,0.000){0}{\usebox{\plotpoint}}
\put(863.42,419.00){\usebox{\plotpoint}}
\multiput(871,419)(20.756,0.000){0}{\usebox{\plotpoint}}
\put(884.18,419.00){\usebox{\plotpoint}}
\put(904.93,419.00){\usebox{\plotpoint}}
\multiput(908,419)(20.756,0.000){0}{\usebox{\plotpoint}}
\put(925.69,419.00){\usebox{\plotpoint}}
\multiput(932,419)(20.756,0.000){0}{\usebox{\plotpoint}}
\put(946.44,419.00){\usebox{\plotpoint}}
\put(967.20,419.00){\usebox{\plotpoint}}
\multiput(969,419)(20.756,0.000){0}{\usebox{\plotpoint}}
\put(987.95,419.00){\usebox{\plotpoint}}
\multiput(994,419)(20.756,0.000){0}{\usebox{\plotpoint}}
\put(1008.71,419.00){\usebox{\plotpoint}}
\put(1029.46,419.00){\usebox{\plotpoint}}
\multiput(1031,419)(20.756,0.000){0}{\usebox{\plotpoint}}
\put(1050.22,419.00){\usebox{\plotpoint}}
\multiput(1055,419)(20.756,0.000){0}{\usebox{\plotpoint}}
\put(1070.98,419.00){\usebox{\plotpoint}}
\put(1091.73,419.00){\usebox{\plotpoint}}
\multiput(1092,419)(20.756,0.000){0}{\usebox{\plotpoint}}
\put(1112.49,419.00){\usebox{\plotpoint}}
\multiput(1117,419)(20.756,0.000){0}{\usebox{\plotpoint}}
\put(1133.24,419.00){\usebox{\plotpoint}}
\multiput(1141,419)(20.756,0.000){0}{\usebox{\plotpoint}}
\put(1154.00,419.00){\usebox{\plotpoint}}
\put(1174.75,419.00){\usebox{\plotpoint}}
\multiput(1178,419)(20.756,0.000){0}{\usebox{\plotpoint}}
\put(1195.51,419.00){\usebox{\plotpoint}}
\multiput(1203,419)(20.756,0.000){0}{\usebox{\plotpoint}}
\put(1216.26,419.00){\usebox{\plotpoint}}
\put(1237.02,419.00){\usebox{\plotpoint}}
\multiput(1239,419)(20.756,0.000){0}{\usebox{\plotpoint}}
\put(1257.77,419.00){\usebox{\plotpoint}}
\multiput(1264,419)(20.756,0.000){0}{\usebox{\plotpoint}}
\put(1278.53,419.00){\usebox{\plotpoint}}
\put(1299.29,419.00){\usebox{\plotpoint}}
\multiput(1301,419)(20.756,0.000){0}{\usebox{\plotpoint}}
\put(1320.04,419.00){\usebox{\plotpoint}}
\multiput(1325,419)(20.756,0.000){0}{\usebox{\plotpoint}}
\put(1340.80,419.00){\usebox{\plotpoint}}
\put(1361.55,419.00){\usebox{\plotpoint}}
\multiput(1362,419)(20.756,0.000){0}{\usebox{\plotpoint}}
\put(1382.31,419.00){\usebox{\plotpoint}}
\multiput(1387,419)(20.756,0.000){0}{\usebox{\plotpoint}}
\put(1403.06,419.00){\usebox{\plotpoint}}
\put(1423.82,419.00){\usebox{\plotpoint}}
\multiput(1424,419)(20.756,0.000){0}{\usebox{\plotpoint}}
\put(1436,419){\usebox{\plotpoint}}
\end{picture}

\caption{Momentum dependence of the correlation function in the presence (solid line) 
and absence (broken line) of a bound state}
\end{figure}
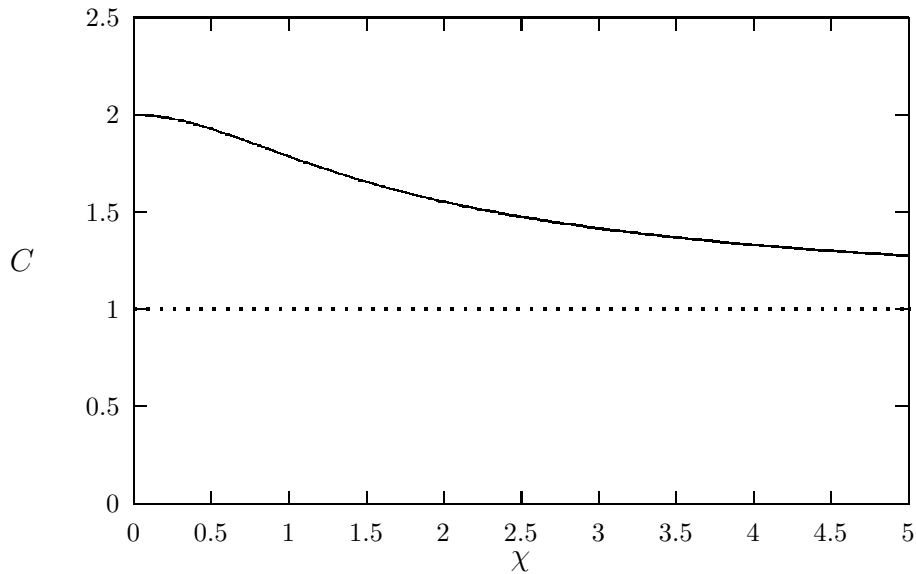
Let us consider it in more detail.
Given a pair of bound $\Delta$-isobars they are localized in a domain with a spatial 
size $\bar{R}\sim 1/\kappa$. For the values of $q$ small compared to 
$1/\bar{R}$ (the ratio $\chi =q/2\kappa \ll 1$) both sources emit the $\pi$-mesons 
coherently resulting in the 100\%-interference and, hence, in $C\rightarrow 2$. In 
the opposite case ($q \gg 1/\bar{R}$) because of the oscillation of the relative
phase (see eq.(\ref{cf})) the interference between the emitted pions is smeared out. 
It results in $C\rightarrow 1$ at large values of $\chi$. The devitation of the 
correlation function $C$ at large $\chi$'s from the unity is proportional to the 
bound state momentum:
\begin{equation}
C-1\approx \frac{\pi}{2}\cdot\frac{2\kappa}{q} 
\end{equation} 

In the case of  unbound $\Delta$-isobars the wave function of their relative
motion (squared) does not have the sense of the probability of finding two
particles at some distance $\vec{R}$. (It is reflected, in particular, in the 
different normalization
of the wave function of the continuous spectrum). This formal obstacle can be
circumvented by considering a continuous-spectrum wave function as that of a
bound state with vanishing binding energy. Then we get  from the
eq.(\ref{ch}) that $C=1$ for all values of $q$. This conclusion follows immediately
from the qualitative consideration. Indeed, since in the continuous spectrum 
arbitrarily large distances between emitting sources are possible the relative phase
controlled by the $\vec{q}\vec{R}$ product will be large for any (nonvanishing)
value of $q$. Inspection of (\ref{cf}) shows that because of rapid oscillations
of the integrand $C=1$ for any unbound state.

When deducing correlation function (\ref{cf}) several approximations have been made. 
The first approximation is that the emission of a $\pi$-meson changes the discrete 
numbers of a baryon $B$  only (e.g. $\Delta\rightarrow N$) and does not affect
its position. Such an assumption seems to be justified in the $N_c\rightarrow\infty$ 
limit where baryons may be considered as (infinitely) heavy particles. Before 
proceeding to the discussion of other approximations made we consider in more detail 
(possible) bound states. The very concept of a bound state of two baryons $B_1$ and 
$B_2$ assumes that they do not lose their identity. It can be translated in the 
requirement that the characteristic distance $\bar{R}$ between two particles exceeds 
their doubled size $b$: $\bar{R}\, \sd \gsim\, \sd 2\,b$. Taking as a benchmark of the
baryon's size that of the nucleon ($b\approx 0.5$~fm) we arrive at the conclusion 
that the spatial dimensions of a two-baryon bound state are to be large: 
$\bar{R}\, \sd \gsim \sd \,1$~fm. The latter inequality translates into the 
limitation on the bound-state momentum $\kappa\approx 1/\bar{R}\sd\lsim\sd 200$~MeV. 
The correlation function $C$  noticeably deviates from 1 for the values of 
$q\approx\kappa$ (see the figure). Strong correlation in the $\pi-\pi$ system are 
known to occur at the effective mass value $\approx 500\div 600$~MeV, i.e. at the 
values of $q$ in the same momentum range. One can conclude that in the $q$ range of
interest ($q\approx 200$~MeV) the correlations between pions are weak thus justifying
the approximation made.

One more implicit assumption made consists in neglecting the $\pi$-meson 
rescattering on the (heavy) particles. Such rescattering could generate 
some phase of the pion wave function (\ref{wf}). The magnitude of the rescattering 
effect can be roughly estimated as follows. Let $\sigma$ be a cross section of the 
emitted $\pi$-meson rescattering on the other baryon. The probability of the process 
is controlled by the ratio $\sigma/(4\pi\bar{R}^2)$. In the considered momentum range 
$|\vec{p}|=|\vec{q}|/2\approx 100\div 200$~MeV the cross section of the $\pi N$
scattering $\sigma_{\pi N}\approx 50\div 100$~mbarn \cite{pdg}. Then the rescattering 
parameter for $\bar{R}\sd\gsim\sd 2$~fm proves to be small $\lsim\sd 25\%$.

Summarizing, possible shallow bound states of two  baryons can be investigated by 
means of the pion interferometry. The momentum dependence  of the two-particle 
correlation function in the "back-to-back" kinematics is controlled by the 
binding-energy value of two baryons. The latter can be determined experimentally 
from the slope of the two-particle correlation function at (small) momenta of 
the $\pi$-mesons. The proposed method suits best for the search of possible 
(loosely) bound states of two resonances which binding, because of the 
large width, can hardly be observed as a shift of a peak in corresponding 
effective-mass distributions.\\[0.15cm]

The author is grateful to M.Polyakov and O.Patarakin for helpful discussions.

\end{document}